\documentclass[aps,prl,twocolumn,showpacs,superscriptaddress]{revtex4-1}
\usepackage{graphicx}

\begin{document}

\title{First-order melting of a weak spin-orbit Mott insulator into a correlated metal}

\author{Tom Hogan}
\affiliation{Department of Physics, Boston College, Chestnut Hill, Massachusetts 02467, USA}
\affiliation{Materials Department, University of California, Santa Barbara, California 93106, USA.}
\author{Z. Yamani}
\affiliation{Canadian Neutron Beam Centre, National Research Council, Chalk River, Ontario, Canada K0J 1P0}
\author{D. Walkup}
\affiliation{Department of Physics, Boston College, Chestnut Hill, Massachusetts 02467, USA}
\author{Xiang Chen}
\affiliation{Department of Physics, Boston College, Chestnut Hill, Massachusetts 02467, USA}
\affiliation{Materials Department, University of California, Santa Barbara, California 93106, USA.}
\author{Rebecca Dally}
\affiliation{Department of Physics, Boston College, Chestnut Hill, Massachusetts 02467, USA}
\affiliation{Materials Department, University of California, Santa Barbara, California 93106, USA.}
\author{Thomas Z. Ward}
\affiliation{Materials Science and Technology Division, Oak Ridge National Laboratory, Oak Ridge, Tennessee 37831, USA}
\author{John Hill}
\affiliation{Department of Condensed Matter Physics and Materials Science, Brookhaven National Laboratory, Upton, New York 11973, USA}
\author{Z. Islam}
\affiliation{The Advanced Photon Source, Argonne National Laboratory, Argonne, Illinois 60439, USA}
\author{Vidya Madhavan}
\affiliation{Department of Physics and Frederick Seitz Materials Research Laboratory,  University of Illinois Urbana-Champaign, Urbana, Illinois  61801, USA.}
\author{Stephen D. Wilson}
\email{stephendwilson@engineering.ucsb.edu}
\affiliation{Materials Department, University of California, Santa Barbara, California 93106, USA.}

\begin{abstract}
The electronic phase diagram of the weak spin-orbit Mott insulator (Sr$_{1-x}$La$_{x}$)$_3$Ir$_2$O$_7$ is determined via an exhaustive experimental study.  Upon doping electrons via La substitution, an immediate collapse in resistivity occurs along with a narrow regime of nanoscale phase separation comprised of antiferromagnetic, insulating regions and paramagnetic, metallic puddles persisting until $x\approx 0.04$.  Continued electron doping results in an abrupt, first-order phase boundary where the N{\'e}el state is suppressed and a homogenous, correlated, metallic state appears with an enhanced spin susceptibility and local moments.  As the metallic state is stabilized, a weak structural distortion develops and suggests a competing instability with the parent spin-orbit Mott state. 
\end{abstract}


\maketitle
The seminal examples of the spin-orbit Mott (SOM) state were reported in the $n=1$ and $n=2$ members of the Sr$_{n+1}$Ir$_{n}$O$_{3n+1}$ Ruddlesden-Popper (RP) series \cite{PhysRevLett.101.076402,PhysRevB.85.184432}, where Ir$^{4+}$ cations, in the limit of a cubic crystal field, realize a $J_{eff}=1/2$ antiferromagnetic (AF) ground state \cite{PhysRevLett112026403,Kim06032009}.  Realizing new electronic phases in close proximity to this SOM state is a subject of considerable theoretical work \cite{Krempa}, and recent experiments have begun to suggest exotic properties present in nearby metallic states \cite{Kim11072014, 2014arXiv1409.8253H}.  However, the central task of understanding the mechanism of the Mott state's collapse in these $5d$-electron Mott systems remains an open question, where, for instance, the roles of competing phases and additional modes of symmetry breaking remain unaddressed.  

The bilayer ($n=2$) material Sr$_3$Ir$_2$O$_7$ (Sr-327) is an excellent test system for exploring carrier substitution in a spin-orbit Mott material \cite{PhysRevB.66.214412,Subramanian1994645}. The reduced short-range Coulomb interaction, $U$, attributable to its $5d$ valence states and the increased bandwidth inherent to Sr-327's bilayer structure lead to a marginally stable insulating state \cite{PhysRevLett.101.226402}.  As a result, the Mott insulating state manifests in the weak limit where the charge gap is of the same order as the nearest neighbor Heisenberg exchange coupling $J$ \cite{PhysRevLett.109.157402, Okada}.  This provides a unique platform for exploring the collapse of the Mott phase, where relatively small perturbations (e.g. changes in carrier concentration) can affect dramatic changes in the stability of the insulating state, and one where the mechanism of the gap's collapse can be explored in the limit of dilute substitution.

Consistent with the idea of a delicate Mott state, Sr-327 has recently been shown to manifest metallic behavior under small levels of La-substitution (electron-doping) \cite{PhysRevB.87.235127}. However, little remains understood regarding the nature of the metallic state realized upon carrier substitution and the means through which the parent $J_{eff}=1/2$ Mott state collapses.  For instance, once the Mott state is destabilized, conflicting reports have suggested both an unusual metallic state with a negative electronic compressibility \cite{2014arXiv1409.8253H} as well as a surprisingly conventional, weakly correlated metal \cite{PhysRevLett.113.256402}.  Notably lacking is a detailed understanding of the structural and electronic responses of this prototypical weak SOM system as electrons are introduced. This remains an essential first step toward developing a deeper understanding of interactions remanent once the parent SOM state is quenched.

Here we present the results of bulk transport/magnetization, neutron/x-ray scattering, and scanning tunneling spectroscopy (STS) measurements mapping the evolution of the antiferromagnetic SOM state in (Sr$_{1-x}$La$_{x}$)$_3$Ir$_2$O$_7$ upon electron substitution.  Light electron doping initially drives the weak SOM state to fragment into nanoscale regions of mixed metallic and insulating character that eventually collapse into a uniform metallic regime beyond $x=0.04$.  The addition of donors to the system causes a swelling of the unit cell volume, and a parallel suppression of magnetostriction effects associated with the onset of Ising-like magnetic order \cite{PhysRevLett.109.037204}.  Once in the globally metallic phase, the long-range G-type N{\'e}el state remnant from the parent Mott phase vanishes, and a metallic state with an enhanced susceptibility and Wilson ratio emerges.  Our aggregate data demonstrate the doping-driven, first-order, melting of a weak spin-orbit Mott phase into a correlated metal.       

Neutron experiments were performed at the N5 triple-axis spectrometer at the Canadian Neutron Beam Centre, Chalk River Laboratories, and resonant x-ray measurements were performed on beam line 6-ID-B at the Advanced Photon Source at Argonne National Lab and X22C at the NSLS at Brookhaven National Lab. Details describing the instrumentation and experimental techniques are provided in the supplemental information \cite{supplemental}.  Crystals were grown via techniques similar to earlier reports \cite{PhysRevB.86.100401, Subramanian1994645}.      

Immediately upon introducing La into Sr-327, a dramatic drop in the low temperature resistivity $\rho (T)$ is observed for concentrations as low as $x=0.01$ as shown in Fig. 1 (a).  Using the na{\"i}ve metric of $\frac{\partial\rho}{\partial T}<0$ as $T\rightarrow 0$ to define an ``insulating" phase, reveals that the system remains in the insulating state until $x_{MIT}\approx0.04$ is reached.  Upon further doping, a change in the sign of the low temperature $\frac{\partial\rho}{\partial T}$ occurs, which we will hereafter denote for simplicity as the metal-insulator transition (MIT).  Doping beyond this level results in the vanishing of the irreversibility in the static spin susceptibility, emblematic of AF ordering in Sr-327, as shown in Fig. 1 (b).  This rapid quenching of the parent system's weak net ferromagnetism is coincident with the onset of the metallic phase and suggests the suppression of the N{\'e}el state in the metallic regime.

As electrons are introduced into the system, the in-plane lattice parameters expand (Fig. 1 (c)) while the c-axis remains unchanged within resolution. This results in a swelling of the lattice volume that continues with increased La-concentration, reminiscent of the lattice swelling observed in La-doped SrTiO$_{3}$ \cite{Janotti} where correlation effects enhance the destabilization/expansion of the lattice driven by adding conduction electrons into antibonding orbitals.   This combined effect competes with steric effects and, if correlation effects are strong enough, can drive lattice expansion even at small doping levels. The relative magnitude of the volumetric expansion $\Delta V/V \approx 0.03\%$ per percentage of La-dopant is nearly identical for both La-doped Sr-327 and La-doped SrTiO$_{3}$---suggesting a comparable role played by correlations. The magnitude of this effect can be demonstrated by alloying comparably-sized, isovalent Ca$^{2+}$ instead of La$^{3+}$ into the system, where purely steric effects instead drive a lattice contraction (Fig 1 (c)).    

\begin{figure}
\includegraphics[scale=.225]{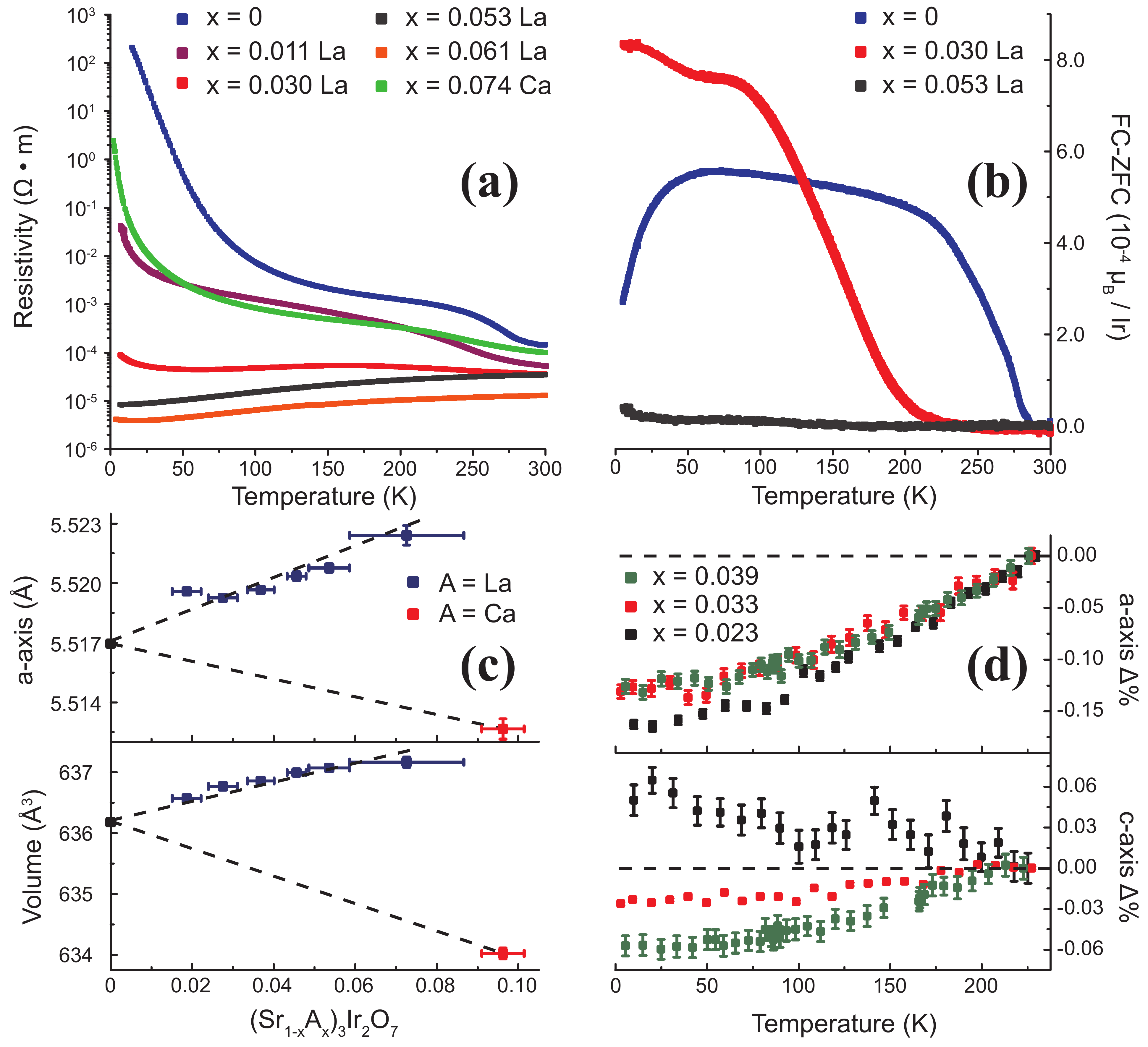}
\caption{(a) Resistivity as a function of temperature $\rho(T)$ for (Sr$_{1-x}$A$_{x}$)$_3$Ir$_2$O$_7$, A=La and Ca.  (b) Magnetization data for La-doped Sr-327.  Data plotted is field-cooled (FC) minus zero-field cooled (ZFC) data under 800 Oe applied parallel to the c-axis. (c) Powder x-ray data showing the $a$-axis and unit cell volume as a function of La- and Ca-substitution in Sr-327. (d) Neutron scattering data showing relative shifts in lattice constants for La-doped Sr-327.  Values are the fractional change of lattice values from 225K, e.g. $a$-axis $\Delta\%=100(\frac{a(T)-a(225K)}{a(225K)})$}
\end{figure}

\begin{figure}
\includegraphics[scale=.225]{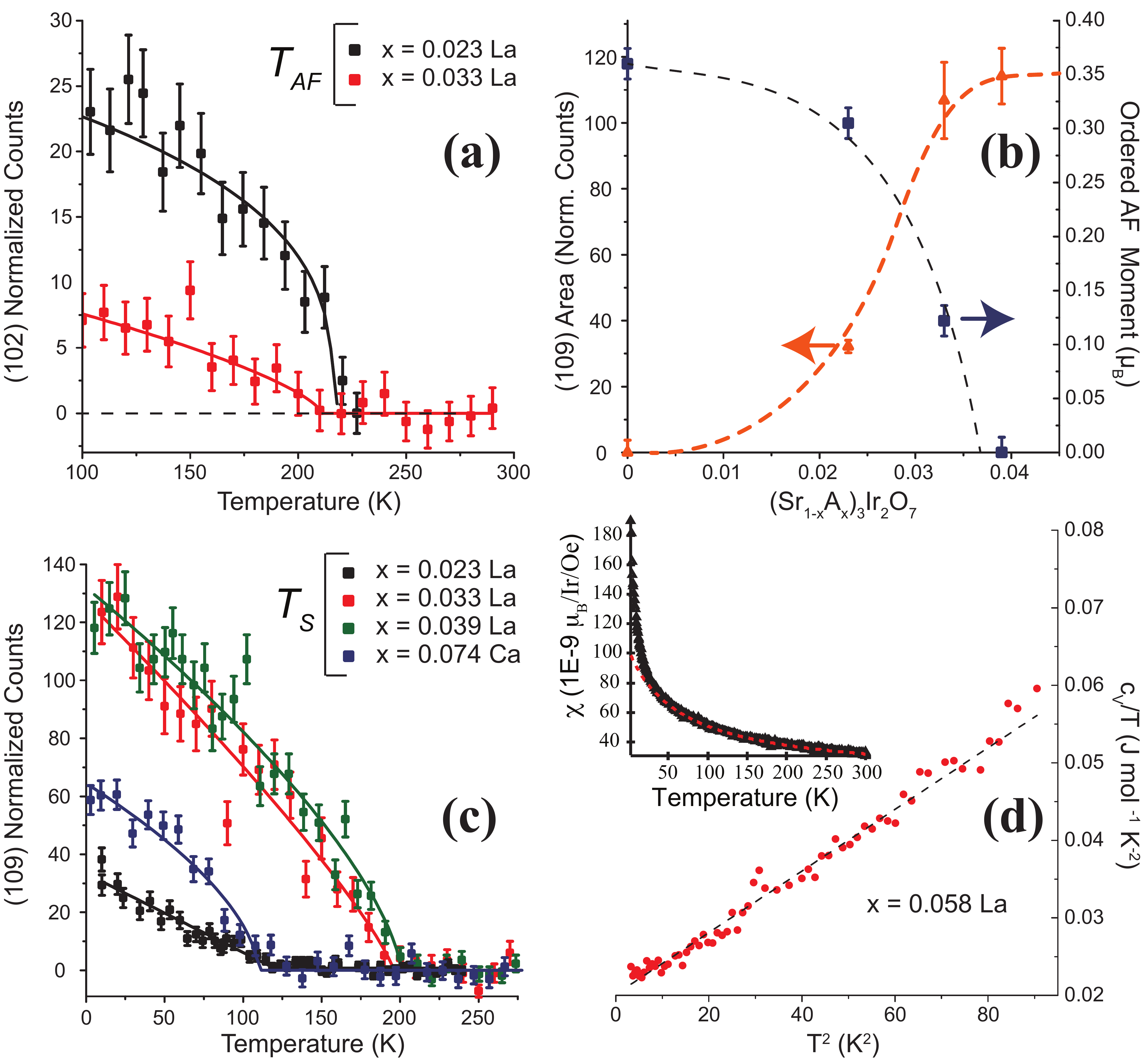}
\caption{(a) Background subtracted magnetic order parameter measurements for La-doped Sr-327.  Data was collected at the \textbf{Q}=$(1,0,2)$ position and normalized to a sample-dependent scale factor. (b) AF-ordered moment and relative weight of forbidden structural peak (1,0,9) (representative of $T_S$ for La-doped Sr-327.  Data for x=0 is taken from \cite{PhysRevB.86.100401}  (c) Background subtracted neutron scattering data showing select $T_S$ order parameters at the (1,0,9) wave vector.  Intensity of the scattering has been normalized via a sample dependent scale factor. (d) Heat capacity $c_v(T)$ data for $x=0.058$ La.  Dashed line is a fit to the form $c_v(T)=\gamma T+\beta T^{3}$ with $\gamma=19.88\pm0.30$ [mJ mole$^{-1}$ K$^{-2}$] and $\beta=0.409\pm0.007$ [mJ mole$^{-1}$ K$^{-4}$].  Inset shows $\chi(T)$ for this same sample with $H=20 kOe\parallel ab$-plane with dotted line denoting the Curie-Weiss fit discussed in the text.  }
\end{figure}

An additional structural response to the MIT is shown in Fig. 1 (d), which reveals that the anisotropic thermal contraction of the parent system upon cooling vanishes as it is doped into the metallic phase.  Namely, both parent and lightly La-doped Sr-327 samples possess a $c$-axis lattice constant that expands upon cooling while the basal plane lattice constants contract. The magnitude of this effect gradually switches to a conventional, uniform, thermal contraction as the MIT is traversed, and the doping-driven switch in behavior tracks the disappearance of irreversibility in the static spin susceptibility. This suggests that the expansion of the c-axis upon cooling for $x\leq x_{MIT}$ is driven by strong magnetoelastic coupling where magnetostriction between the Ising-like, $c$-axis oriented, moments and their local lattice environment drive an anisotropic distortion of the lattice.  

The disappearance of irreversibility in magnetization measurements, however, is not a rigorous metric for determining the doping evolution of the magnetic order in a canted AF.  To further investigate the evolution of AF order as the metallic state is approached, neutron scattering measurements were performed.  For samples with $x\leq x_{MIT}$, magnetic scattering remained consistent with the G-type spin structure of the parent material \cite{PhysRevB.86.100401, PhysRevLett.109.037204}.  Scattering results plotted in Figs. 2 (a) and (b) show that the ordered AF moment rapidly collapses as $x_{MIT}$ is approached, yet the ordering temperature remains only weakly affected.  This contrasts the percolative MIT realized in Ru-doped Sr-327, where AF order survives into the metallic regime and remains coherent across electronically phase separated patches \cite{DhitalRuDoped}.  Instead, La-substitution rapidly quenches spin order associated with Sr-327's G-type structure, which vanishes with the stabilization of the low temperature metallic state.     

\begin{figure}
\includegraphics[scale=.55]{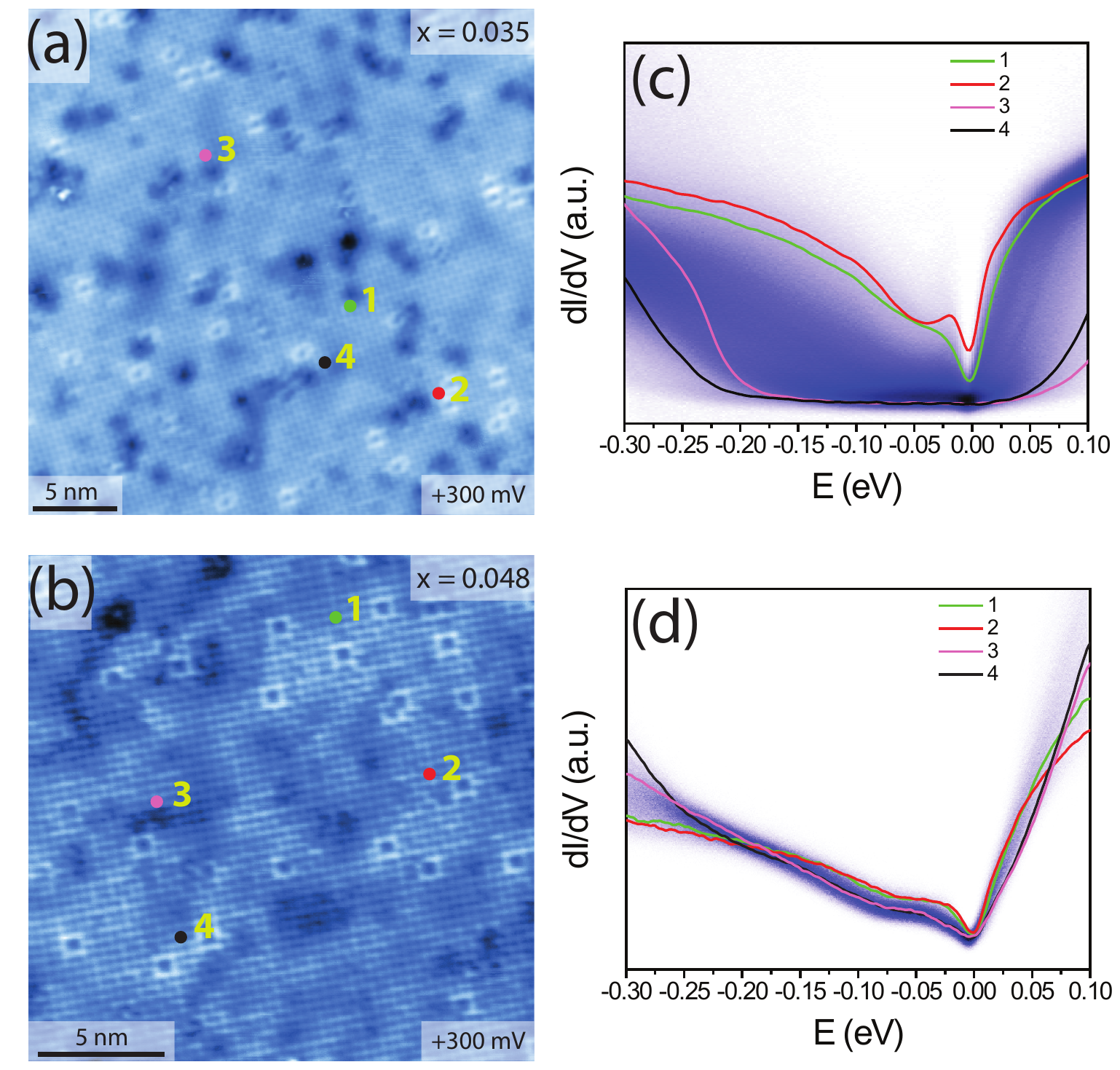}
\caption{STM topography of La-doped Sr-327 at 300 mV bias for (a) insulating x=0.035 and (b) metallic x=0.048 samples.  $\frac{dI}{dV}$ spectra obtained on a grid in each topograph are plotted in panels (c) for x=0.035 and (d) for x=0.048.  Representative numbered points are highlighted in each map and the corresponding $\frac{dI}{dV}$ spectra are emphasized as solid lines in spectral histograms.  Fully gapped and gapless spectra are observed in the x=0.035 sample while a homogenous, gapless state appears across spectra in x=0.048.}
\end{figure}

An additional structural order parameter also develops as a function of La-doping below a characteristic temperature $T_S$.  This distortion appears in the form of a weak, temperature dependent, structural superlattice at \textit{Bbcb} forbidden $(H = odd, 0, L = odd)$ positions.  Figs. 2 (b-c) show the evolution of this distortion as a function of increasing La-content.  The relative weights of Bragg reflections associated with this distortion are plotted in Fig. 2 (b), and the corresponding temperature evolution of the order parameters are plotted in Fig. 2 (c).  As La content is increased, both $T_S$ and its relative scattering weight increase, seemingly saturating across $x_{MIT}$.  

One explanation for the trade-off in scattering weight between this new structural order parameter and AF order, along with the relatively weak doping dependence of $T_S$ and $T_{AF}$ away from the critical regime, is that light electron-doping generates a phase separated ground state.  To test this notion, STS measurements were performed on samples residing on both sides of the MIT. The resulting spectra of samples in the insulating $x=0.035$ and metallic $x=0.048$ regimes are plotted in Fig. 3 where electron-doping with $x\leq x_{MIT}$ results in a nanoscale phase separated ground state with distinct insulating and gapless regions.  Upon continued doping to $x=0.048$, a homogenous, globally gapless, ground state is observed and is consistent with the metallic transport observed for $x>x_{MIT}$.   

Beyond $x_{MIT}$, static spin susceptibility data for a metallic sample with $x=0.058$ are plotted in Fig. 2 (d) inset.  The data fit to a Curie-Weiss model with an additional temperature independent Pauli term, giving $\Theta=-69\pm9$ K and $\mu_{eff}=0.51\pm0.02$ $\mu_B$.  The potential of this local moment response arising from trivial inhomogeneity (ie. rare regions of this sample with clustered spins and a persistent charge gap) can be excluded via comparison with the globally gapless STS data in Fig. 3 (b) and (d) \cite{supplemental}.  Assuming that the surface electronic states probed by STS data are reflective of the bulk, the combined analysis of the susceptibility and STS data mandates the survival of a local moment response within gapless regions of the sample.     

Heat capacity data from this same $x=0.058$ concentration (Fig. 2 (d)) obtain a Sommerfeld coefficient $\gamma=19.88\pm0.30$ [mJ mole$^{-1}$ K$^{-2}$] ($\gamma=9.94$ [mJ mole-Ir$^{-1}$ K$^{-2}$]), also reflecting a metal with enhanced correlation effects.  Low temperature $\chi (T)$ from this same sample shows $\chi=0.0229$ [J T$^{-2}$ mole$^{-1}$] at $T=2$ K, leading to a Wilson ratio of $R_{W}=\frac{\pi^2k_B^2\chi}{3\mu_B^2\gamma}\approx 8.4$.  This enhanced $R_W$ is consistent with a system near an instability \cite{PhysRevB.62.R6089} and suggests that the state realized for $x>x_{MIT}$ is a correlated metal with an enhanced spin susceptibility; retaining remnant correlations from the SOM parent phase.  

The electronic phase diagram summarizing the evolution of the SOM phase upon electron-doping is plotted in Fig. 4.  Immediately upon doping electrons into the parent Sr-327, a regime of phase separation appears---one where nanoscale AF ordered insulating regions segregate from gapless metallic regions that stabilize a global structural distortion below $T_S$.  For $x<x_{MIT}$, $T_S$ increases in parallel to the growth of the volume fraction of the sample hosting the metallic phase. Similarly, the combined neutron/STM data of Figs. 2 and 3 demonstrate that the apparent reduction in the AF moment under light electron doping largely arises from electronic phase separation of the sample into AF ordered insulating and paramagnetic metallic regions. Upon doping beyond the critical concentration of $x\approx0.04$, a first-order line appears where AF order collapses and the system becomes globally metallic.       

Earlier reports of persistent AF order in metallic concentrations of La-doped Sr-327 were unable to discern whether this coexistence was intrinsic to the physics of a doped SOM insulator or extrinsic due to macroscopic sample inhomogeneity \cite{PhysRevB.87.235127}.  Our observation of an abrupt, first-order, collapse of the Sr-327 parent material's N{\'e}el state upon entering the metallic regime resolves this open question and demonstrates the instability of long-range AF order once the weak SOM state inherent to Sr-327 is tuned beyond half-filling.  Since strong in-plane AF superexchange masks the local moment behavior above $T_{AF}$ in undoped Sr-327 \cite{PhysRevLett.109.157402, Nagai}, the doping-induced collapse of AF beyond the MIT ultimately allows for the Ir local moments to be observed in the correlated metallic regime.  
 
\begin{figure}
\includegraphics[scale=.65]{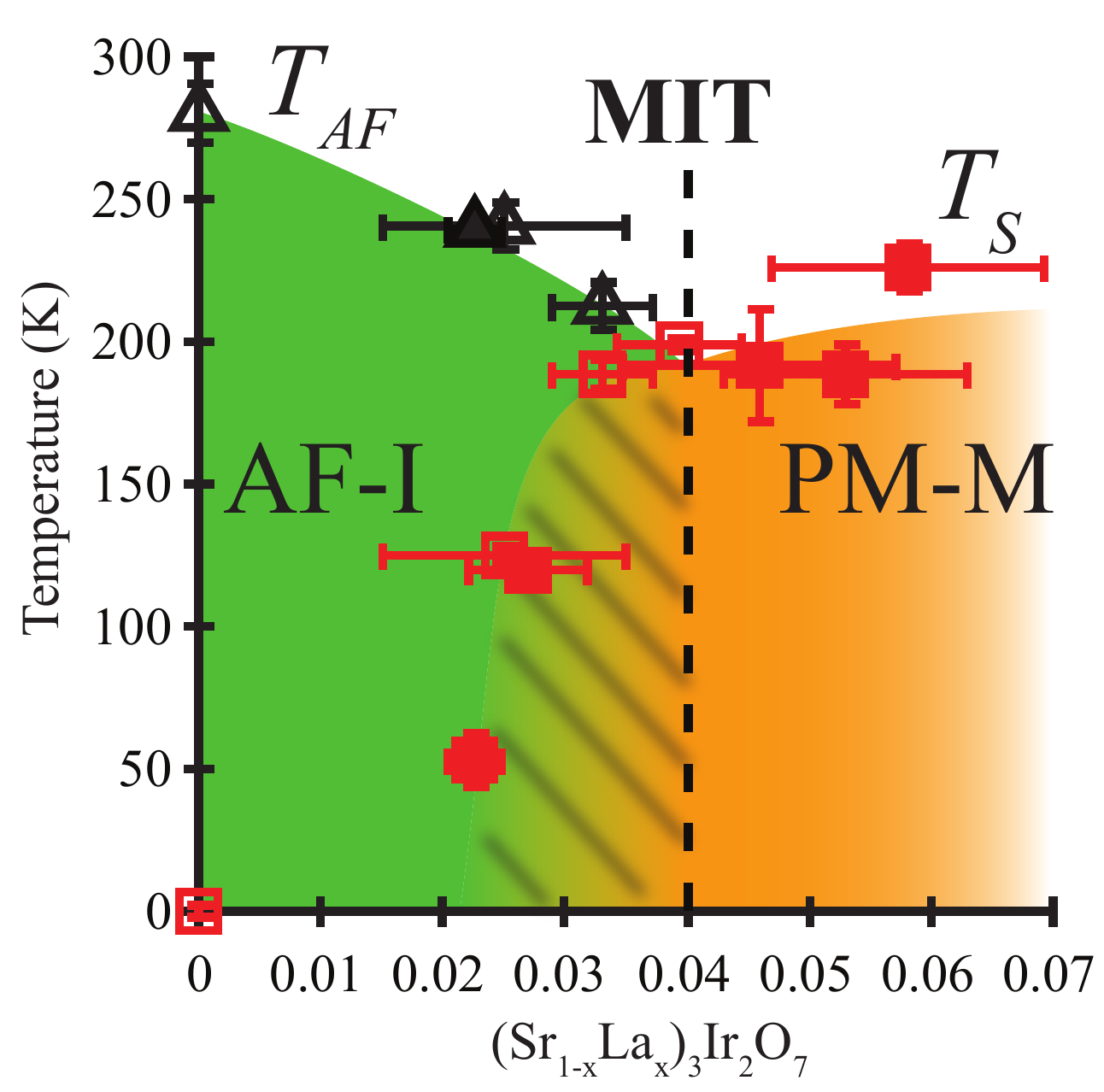}
\caption{Electronic phase diagram of La-doped Sr-327.  Open symbols denote neutron scattering measurements and closed symbols denote x-ray data.  Black triangles denote scattering measurements of $T_{AF}$ and red squares denote measurements of $T_S$.  The dashed line marks the MIT measured via charge transport and the first-order line where the N{\'e}el state vanishes.  The hatched region marks the electronically phase separated region where $T_S$ and $T_{AF}$, AF-I denotes antiferromagnetic insulating state, and PM-M marks the paramagnetic metallic phase.  Data for x=0 taken from \cite{PhysRevB.86.100401}} 
\end{figure}

Our measurements depict the destruction of the parent state's N{\'e}el order upon entering the metallic regime and are consistent with recent theoretical work demonstrating the filling-tuned, first-order MIT of a weak Mott state in the intermediate coupling regime \cite{PhysRevX.5.021007, 0295-5075-85-1-17002}.  The first order nature of the MIT is demonstrated explicitly by the phase coexistence for $x\leq x_{MIT}$ plotted in Fig. 3.  As the system is driven across the MIT phase boundary, the development of a structural symmetry breaking transition suggests a multicritical point driven by a competing energy scale, $T_S$, near the parent SOM phase.  

As one test of whether $T_S$ is endemic to the metallic state, neutron scattering measurements were performed on isovalent-substituted (Sr$_{0.93}$Ca$_{0.07}$)$_3$Ir$_2$O$_7$.  This system remains an insulator (Fig. 1 (a)), yet the reduced cation size drives a low temperature structural distortion along the identical $(odd, 0, odd)$ wave vectors as La-substituted Sr-327  (Fig. 2 (c)). $T_S$ for this Ca-doped sample, however, occurs at a reduced energy scale relative to its La-doped counterpart. This reduced $T_S$ at a comparatively higher Ca-dopant concentration (larger steric perturbation) suggests that the electronic contribution to the lattice deformation enhances $T_S$ and drives the metallic state.  This is also consistent with recent reports of a structural distortion appearing near the pressure-driven MIT of Sr-327 \cite{Zhaopressure, PhysRevB.87.235127}.  

In summary, our data demonstrate the carrier-driven first-order melting of the SOM phase in (Sr$_{1-x}$La$_{x}$)$_3$Ir$_2$O$_7$, consistent with the predictions of an electronically phase separated state intermediate to the complete collapse of the Mott phase in the weak limit.  Beyond the critical $x_{MIT}=0.04$ concentration, the SOM state collapses into a metallic state with enhanced spin susceptibility and local moment behavior.  Ascertaining whether the lattice distortion emergent at the onset of the metallic state is purely a structural effect or a secondary consequence of a competing electronic instability is an interesting avenue for future exploration.  

\acknowledgments{
S.D.W. thanks L. Balents, R. Seshadri, and Z. Wang for helpful discussions.  This work was supported in part by NSF CAREER award DMR-1056625 (S.D.W.). T. Hogan and magnetization measurements were partially supported by DMR-1121053.  Partial support given by the US Department of Energy (DOE), Office of Basic Energy Sciences (BES), Materials Sciences and Engineering Division, (T.Z.W.) STM work (V.M. and D.W.) were supported by grant from the National Science Foundation, NSF DMR-1305647.  The work at the Advanced Photon Source of Argonne National Laboratory was supported by the U.S. Department of Energy Basic Energy Sciences under Contract No. NE-AC02-06CH11357. Work at Brookhaven National Laboratory and the National Synchrotron Light Source,was supported by the U.S. Department of Energy, Office of Science, Office of Basic Energy Sciences, under Contract No. DE-AC02-98CH10886}

\bibliography{LaDopedBib}

\begin{thebibliography}{24}
\expandafter\ifx\csname natexlab\endcsname\relax\def\natexlab#1{#1}\fi
\expandafter\ifx\csname bibnamefont\endcsname\relax
  \def\bibnamefont#1{#1}\fi
\expandafter\ifx\csname bibfnamefont\endcsname\relax
  \def\bibfnamefont#1{#1}\fi
\expandafter\ifx\csname citenamefont\endcsname\relax
  \def\citenamefont#1{#1}\fi
\expandafter\ifx\csname url\endcsname\relax
  \def\url#1{\texttt{#1}}\fi
\expandafter\ifx\csname urlprefix\endcsname\relax\def\urlprefix{URL }\fi
\providecommand{\bibinfo}[2]{#2}
\providecommand{\eprint}[2][]{\url{#2}}

\bibitem[{\citenamefont{Kim et~al.}(2008)\citenamefont{Kim, Jin, Moon, Kim,
  Park, Leem, Yu, Noh, Kim, Oh et~al.}}]{PhysRevLett.101.076402}
\bibinfo{author}{\bibfnamefont{B.~J.} \bibnamefont{Kim}},
  \bibinfo{author}{\bibfnamefont{H.}~\bibnamefont{Jin}},
  \bibinfo{author}{\bibfnamefont{S.~J.} \bibnamefont{Moon}},
  \bibinfo{author}{\bibfnamefont{J.-Y.} \bibnamefont{Kim}},
  \bibinfo{author}{\bibfnamefont{B.-G.} \bibnamefont{Park}},
  \bibinfo{author}{\bibfnamefont{C.~S.} \bibnamefont{Leem}},
  \bibinfo{author}{\bibfnamefont{J.}~\bibnamefont{Yu}},
  \bibinfo{author}{\bibfnamefont{T.~W.} \bibnamefont{Noh}},
  \bibinfo{author}{\bibfnamefont{C.}~\bibnamefont{Kim}},
  \bibinfo{author}{\bibfnamefont{S.-J.} \bibnamefont{Oh}},
  \bibnamefont{et~al.}, \bibinfo{journal}{Phys. Rev. Lett.}
  \textbf{\bibinfo{volume}{101}}, \bibinfo{pages}{076402}
  (\bibinfo{year}{2008}).

\bibitem[{\citenamefont{Boseggia et~al.}(2012)\citenamefont{Boseggia,
  Springell, Walker, Boothroyd, Prabhakaran, Wermeille, Bouchenoire, Collins,
  and McMorrow}}]{PhysRevB.85.184432}
\bibinfo{author}{\bibfnamefont{S.}~\bibnamefont{Boseggia}},
  \bibinfo{author}{\bibfnamefont{R.}~\bibnamefont{Springell}},
  \bibinfo{author}{\bibfnamefont{H.~C.} \bibnamefont{Walker}},
  \bibinfo{author}{\bibfnamefont{A.~T.} \bibnamefont{Boothroyd}},
  \bibinfo{author}{\bibfnamefont{D.}~\bibnamefont{Prabhakaran}},
  \bibinfo{author}{\bibfnamefont{D.}~\bibnamefont{Wermeille}},
  \bibinfo{author}{\bibfnamefont{L.}~\bibnamefont{Bouchenoire}},
  \bibinfo{author}{\bibfnamefont{S.~P.} \bibnamefont{Collins}},
  \bibnamefont{and} \bibinfo{author}{\bibfnamefont{D.~F.}
  \bibnamefont{McMorrow}}, \bibinfo{journal}{Phys. Rev. B}
  \textbf{\bibinfo{volume}{85}}, \bibinfo{pages}{184432}
  (\bibinfo{year}{2012}).

\bibitem[{\citenamefont{Moretti~Sala et~al.}(2014)\citenamefont{Moretti~Sala,
  Boseggia, McMorrow, and Monaco}}]{PhysRevLett112026403}
\bibinfo{author}{\bibfnamefont{M.}~\bibnamefont{Moretti~Sala}},
  \bibinfo{author}{\bibfnamefont{S.}~\bibnamefont{Boseggia}},
  \bibinfo{author}{\bibfnamefont{F.}~\bibnamefont{McMorrow},
  \bibfnamefont{D.}}, \bibnamefont{and}
  \bibinfo{author}{\bibfnamefont{G.}~\bibnamefont{Monaco}},
  \bibinfo{journal}{Phys. Rev. Lett.} \textbf{\bibinfo{volume}{112}},
  \bibinfo{pages}{026403} (\bibinfo{year}{2014}).

\bibitem[{\citenamefont{Kim et~al.}(2009)\citenamefont{Kim, Ohsumi, Komesu,
  Sakai, Morita, Takagi, and Arima}}]{Kim06032009}
\bibinfo{author}{\bibfnamefont{B.~J.} \bibnamefont{Kim}},
  \bibinfo{author}{\bibfnamefont{H.}~\bibnamefont{Ohsumi}},
  \bibinfo{author}{\bibfnamefont{T.}~\bibnamefont{Komesu}},
  \bibinfo{author}{\bibfnamefont{S.}~\bibnamefont{Sakai}},
  \bibinfo{author}{\bibfnamefont{T.}~\bibnamefont{Morita}},
  \bibinfo{author}{\bibfnamefont{H.}~\bibnamefont{Takagi}}, \bibnamefont{and}
  \bibinfo{author}{\bibfnamefont{T.}~\bibnamefont{Arima}},
  \bibinfo{journal}{Science} \textbf{\bibinfo{volume}{323}},
  \bibinfo{pages}{1329} (\bibinfo{year}{2009}).

\bibitem[{\citenamefont{Witczak-Krempa
  et~al.}(2014)\citenamefont{Witczak-Krempa, Chen, Kim, and Balents}}]{Krempa}
\bibinfo{author}{\bibfnamefont{W.}~\bibnamefont{Witczak-Krempa}},
  \bibinfo{author}{\bibfnamefont{G.}~\bibnamefont{Chen}},
  \bibinfo{author}{\bibfnamefont{Y.~B.} \bibnamefont{Kim}}, \bibnamefont{and}
  \bibinfo{author}{\bibfnamefont{L.}~\bibnamefont{Balents}},
  \bibinfo{journal}{Annual Review of Condensed Matter Physics}
  \textbf{\bibinfo{volume}{5}}, \bibinfo{pages}{57} (\bibinfo{year}{2014}).

\bibitem[{\citenamefont{Kim et~al.}(2014)\citenamefont{Kim, Krupin, Denlinger,
  Bostwick, Rotenberg, Zhao, Mitchell, Allen, and Kim}}]{Kim11072014}
\bibinfo{author}{\bibfnamefont{Y.~K.} \bibnamefont{Kim}},
  \bibinfo{author}{\bibfnamefont{O.}~\bibnamefont{Krupin}},
  \bibinfo{author}{\bibfnamefont{J.~D.} \bibnamefont{Denlinger}},
  \bibinfo{author}{\bibfnamefont{A.}~\bibnamefont{Bostwick}},
  \bibinfo{author}{\bibfnamefont{E.}~\bibnamefont{Rotenberg}},
  \bibinfo{author}{\bibfnamefont{Q.}~\bibnamefont{Zhao}},
  \bibinfo{author}{\bibfnamefont{J.~F.} \bibnamefont{Mitchell}},
  \bibinfo{author}{\bibfnamefont{J.~W.} \bibnamefont{Allen}}, \bibnamefont{and}
  \bibinfo{author}{\bibfnamefont{B.~J.} \bibnamefont{Kim}},
  \bibinfo{journal}{Science} \textbf{\bibinfo{volume}{345}},
  \bibinfo{pages}{187} (\bibinfo{year}{2014}).

\bibitem[{\citenamefont{{He} et~al.}(2014)\citenamefont{{He}, {Hogan}, {Mion},
  {Hafiz}, {He}, {Mo}, {Dhital}, {Chen}, {Lin}, {Zhang}
  et~al.}}]{2014arXiv1409.8253H}
\bibinfo{author}{\bibfnamefont{J.}~\bibnamefont{{He}}},
  \bibinfo{author}{\bibfnamefont{T.}~\bibnamefont{{Hogan}}},
  \bibinfo{author}{\bibfnamefont{T.~R.} \bibnamefont{{Mion}}},
  \bibinfo{author}{\bibfnamefont{H.}~\bibnamefont{{Hafiz}}},
  \bibinfo{author}{\bibfnamefont{Y.}~\bibnamefont{{He}}},
  \bibinfo{author}{\bibfnamefont{S.-K.} \bibnamefont{{Mo}}},
  \bibinfo{author}{\bibfnamefont{C.}~\bibnamefont{{Dhital}}},
  \bibinfo{author}{\bibfnamefont{X.}~\bibnamefont{{Chen}}},
  \bibinfo{author}{\bibfnamefont{Q.}~\bibnamefont{{Lin}}},
  \bibinfo{author}{\bibfnamefont{Y.}~\bibnamefont{{Zhang}}},
  \bibnamefont{et~al.}, \bibinfo{journal}{ArXiv e-prints}
  (\bibinfo{year}{2014}), \eprint{1409.8253}.

\bibitem[{\citenamefont{Cao et~al.}(2002)\citenamefont{Cao, Xin, Alexander,
  Crow, Schlottmann, Crawford, Harlow, and Marshall}}]{PhysRevB.66.214412}
\bibinfo{author}{\bibfnamefont{G.}~\bibnamefont{Cao}},
  \bibinfo{author}{\bibfnamefont{Y.}~\bibnamefont{Xin}},
  \bibinfo{author}{\bibfnamefont{C.~S.} \bibnamefont{Alexander}},
  \bibinfo{author}{\bibfnamefont{J.~E.} \bibnamefont{Crow}},
  \bibinfo{author}{\bibfnamefont{P.}~\bibnamefont{Schlottmann}},
  \bibinfo{author}{\bibfnamefont{M.~K.} \bibnamefont{Crawford}},
  \bibinfo{author}{\bibfnamefont{R.~L.} \bibnamefont{Harlow}},
  \bibnamefont{and} \bibinfo{author}{\bibfnamefont{W.}~\bibnamefont{Marshall}},
  \bibinfo{journal}{Phys. Rev. B} \textbf{\bibinfo{volume}{66}},
  \bibinfo{pages}{214412} (\bibinfo{year}{2002}).

\bibitem[{\citenamefont{Subramanian et~al.}(1994)\citenamefont{Subramanian,
  Crawford, and Harlow}}]{Subramanian1994645}
\bibinfo{author}{\bibfnamefont{M.}~\bibnamefont{Subramanian}},
  \bibinfo{author}{\bibfnamefont{M.}~\bibnamefont{Crawford}}, \bibnamefont{and}
  \bibinfo{author}{\bibfnamefont{R.}~\bibnamefont{Harlow}},
  \bibinfo{journal}{Materials Research Bulletin} \textbf{\bibinfo{volume}{29}},
  \bibinfo{pages}{645 } (\bibinfo{year}{1994}).

\bibitem[{\citenamefont{Moon et~al.}(2008)\citenamefont{Moon, Jin, Kim, Choi,
  Lee, Yu, Cao, Sumi, Funakubo, Bernhard et~al.}}]{PhysRevLett.101.226402}
\bibinfo{author}{\bibfnamefont{S.~J.} \bibnamefont{Moon}},
  \bibinfo{author}{\bibfnamefont{H.}~\bibnamefont{Jin}},
  \bibinfo{author}{\bibfnamefont{K.~W.} \bibnamefont{Kim}},
  \bibinfo{author}{\bibfnamefont{W.~S.} \bibnamefont{Choi}},
  \bibinfo{author}{\bibfnamefont{Y.~S.} \bibnamefont{Lee}},
  \bibinfo{author}{\bibfnamefont{J.}~\bibnamefont{Yu}},
  \bibinfo{author}{\bibfnamefont{G.}~\bibnamefont{Cao}},
  \bibinfo{author}{\bibfnamefont{A.}~\bibnamefont{Sumi}},
  \bibinfo{author}{\bibfnamefont{H.}~\bibnamefont{Funakubo}},
  \bibinfo{author}{\bibfnamefont{C.}~\bibnamefont{Bernhard}},
  \bibnamefont{et~al.}, \bibinfo{journal}{Phys. Rev. Lett.}
  \textbf{\bibinfo{volume}{101}}, \bibinfo{pages}{226402}
  (\bibinfo{year}{2008}).

\bibitem[{\citenamefont{Kim et~al.}(2012{\natexlab{a}})\citenamefont{Kim, Said,
  Casa, Upton, Gog, Daghofer, Jackeli, van~den Brink, Khaliullin, and
  Kim}}]{PhysRevLett.109.157402}
\bibinfo{author}{\bibfnamefont{J.}~\bibnamefont{Kim}},
  \bibinfo{author}{\bibfnamefont{A.~H.} \bibnamefont{Said}},
  \bibinfo{author}{\bibfnamefont{D.}~\bibnamefont{Casa}},
  \bibinfo{author}{\bibfnamefont{M.~H.} \bibnamefont{Upton}},
  \bibinfo{author}{\bibfnamefont{T.}~\bibnamefont{Gog}},
  \bibinfo{author}{\bibfnamefont{M.}~\bibnamefont{Daghofer}},
  \bibinfo{author}{\bibfnamefont{G.}~\bibnamefont{Jackeli}},
  \bibinfo{author}{\bibfnamefont{J.}~\bibnamefont{van~den Brink}},
  \bibinfo{author}{\bibfnamefont{G.}~\bibnamefont{Khaliullin}},
  \bibnamefont{and} \bibinfo{author}{\bibfnamefont{B.~J.} \bibnamefont{Kim}},
  \bibinfo{journal}{Phys. Rev. Lett.} \textbf{\bibinfo{volume}{109}},
  \bibinfo{pages}{157402} (\bibinfo{year}{2012}{\natexlab{a}}).

\bibitem[{\citenamefont{Okada et~al.}(2013)\citenamefont{Okada, Walkup, Lin,
  Dhital, Chang, Khadka, Zhou, Jeng, Paranjape, Bansil et~al.}}]{Okada}
\bibinfo{author}{\bibfnamefont{Y.}~\bibnamefont{Okada}},
  \bibinfo{author}{\bibfnamefont{D.}~\bibnamefont{Walkup}},
  \bibinfo{author}{\bibfnamefont{H.}~\bibnamefont{Lin}},
  \bibinfo{author}{\bibfnamefont{C.}~\bibnamefont{Dhital}},
  \bibinfo{author}{\bibfnamefont{T.-R.} \bibnamefont{Chang}},
  \bibinfo{author}{\bibfnamefont{S.}~\bibnamefont{Khadka}},
  \bibinfo{author}{\bibfnamefont{W.}~\bibnamefont{Zhou}},
  \bibinfo{author}{\bibfnamefont{H.-T.} \bibnamefont{Jeng}},
  \bibinfo{author}{\bibfnamefont{M.}~\bibnamefont{Paranjape}},
  \bibinfo{author}{\bibfnamefont{A.}~\bibnamefont{Bansil}},
  \bibnamefont{et~al.}, \bibinfo{journal}{Nat Mater}
  \textbf{\bibinfo{volume}{12}}, \bibinfo{pages}{707} (\bibinfo{year}{2013}).

\bibitem[{\citenamefont{Li et~al.}(2013)\citenamefont{Li, Kong, Qi, Jin, Yuan,
  DeLong, Schlottmann, and Cao}}]{PhysRevB.87.235127}
\bibinfo{author}{\bibfnamefont{L.}~\bibnamefont{Li}},
  \bibinfo{author}{\bibfnamefont{P.~P.} \bibnamefont{Kong}},
  \bibinfo{author}{\bibfnamefont{T.~F.} \bibnamefont{Qi}},
  \bibinfo{author}{\bibfnamefont{C.~Q.} \bibnamefont{Jin}},
  \bibinfo{author}{\bibfnamefont{S.~J.} \bibnamefont{Yuan}},
  \bibinfo{author}{\bibfnamefont{L.~E.} \bibnamefont{DeLong}},
  \bibinfo{author}{\bibfnamefont{P.}~\bibnamefont{Schlottmann}},
  \bibnamefont{and} \bibinfo{author}{\bibfnamefont{G.}~\bibnamefont{Cao}},
  \bibinfo{journal}{Phys. Rev. B} \textbf{\bibinfo{volume}{87}},
  \bibinfo{pages}{235127} (\bibinfo{year}{2013}).

\bibitem[{\citenamefont{de~la Torre et~al.}(2014)\citenamefont{de~la Torre,
  Hunter, Subedi, McKeown~Walker, Tamai, Kim, Hoesch, Perry, Georges, and
  Baumberger}}]{PhysRevLett.113.256402}
\bibinfo{author}{\bibfnamefont{A.}~\bibnamefont{de~la Torre}},
  \bibinfo{author}{\bibfnamefont{E.~C.} \bibnamefont{Hunter}},
  \bibinfo{author}{\bibfnamefont{A.}~\bibnamefont{Subedi}},
  \bibinfo{author}{\bibfnamefont{S.}~\bibnamefont{McKeown~Walker}},
  \bibinfo{author}{\bibfnamefont{A.}~\bibnamefont{Tamai}},
  \bibinfo{author}{\bibfnamefont{T.~K.} \bibnamefont{Kim}},
  \bibinfo{author}{\bibfnamefont{M.}~\bibnamefont{Hoesch}},
  \bibinfo{author}{\bibfnamefont{R.~S.} \bibnamefont{Perry}},
  \bibinfo{author}{\bibfnamefont{A.}~\bibnamefont{Georges}}, \bibnamefont{and}
  \bibinfo{author}{\bibfnamefont{F.}~\bibnamefont{Baumberger}},
  \bibinfo{journal}{Phys. Rev. Lett.} \textbf{\bibinfo{volume}{113}},
  \bibinfo{pages}{256402} (\bibinfo{year}{2014}).

\bibitem[{\citenamefont{Kim et~al.}(2012{\natexlab{b}})\citenamefont{Kim, Choi,
  Kim, Mitchell, Jackeli, Daghofer, van~den Brink, Khaliullin, and
  Kim}}]{PhysRevLett.109.037204}
\bibinfo{author}{\bibfnamefont{J.~W.} \bibnamefont{Kim}},
  \bibinfo{author}{\bibfnamefont{Y.}~\bibnamefont{Choi}},
  \bibinfo{author}{\bibfnamefont{J.}~\bibnamefont{Kim}},
  \bibinfo{author}{\bibfnamefont{J.~F.} \bibnamefont{Mitchell}},
  \bibinfo{author}{\bibfnamefont{G.}~\bibnamefont{Jackeli}},
  \bibinfo{author}{\bibfnamefont{M.}~\bibnamefont{Daghofer}},
  \bibinfo{author}{\bibfnamefont{J.}~\bibnamefont{van~den Brink}},
  \bibinfo{author}{\bibfnamefont{G.}~\bibnamefont{Khaliullin}},
  \bibnamefont{and} \bibinfo{author}{\bibfnamefont{B.~J.} \bibnamefont{Kim}},
  \bibinfo{journal}{Phys. Rev. Lett.} \textbf{\bibinfo{volume}{109}},
  \bibinfo{pages}{037204} (\bibinfo{year}{2012}{\natexlab{b}}).

\bibitem[{sup()}]{supplemental}
\bibinfo{note}{See Supplemental Information for further details}.

\bibitem[{\citenamefont{Dhital et~al.}(2012)\citenamefont{Dhital, Khadka,
  Yamani, de~la Cruz, Hogan, Disseler, Pokharel, Lukas, Tian, Opeil
  et~al.}}]{PhysRevB.86.100401}
\bibinfo{author}{\bibfnamefont{C.}~\bibnamefont{Dhital}},
  \bibinfo{author}{\bibfnamefont{S.}~\bibnamefont{Khadka}},
  \bibinfo{author}{\bibfnamefont{Z.}~\bibnamefont{Yamani}},
  \bibinfo{author}{\bibfnamefont{C.}~\bibnamefont{de~la Cruz}},
  \bibinfo{author}{\bibfnamefont{T.~C.} \bibnamefont{Hogan}},
  \bibinfo{author}{\bibfnamefont{S.~M.} \bibnamefont{Disseler}},
  \bibinfo{author}{\bibfnamefont{M.}~\bibnamefont{Pokharel}},
  \bibinfo{author}{\bibfnamefont{K.~C.} \bibnamefont{Lukas}},
  \bibinfo{author}{\bibfnamefont{W.}~\bibnamefont{Tian}},
  \bibinfo{author}{\bibfnamefont{C.~P.} \bibnamefont{Opeil}},
  \bibnamefont{et~al.}, \bibinfo{journal}{Phys. Rev. B}
  \textbf{\bibinfo{volume}{86}}, \bibinfo{pages}{100401}
  (\bibinfo{year}{2012}).

\bibitem[{\citenamefont{Janotti et~al.}(2012)\citenamefont{Janotti, Jalan,
  Stemmer, and Van~de Walle}}]{Janotti}
\bibinfo{author}{\bibfnamefont{A.}~\bibnamefont{Janotti}},
  \bibinfo{author}{\bibfnamefont{B.}~\bibnamefont{Jalan}},
  \bibinfo{author}{\bibfnamefont{S.}~\bibnamefont{Stemmer}}, \bibnamefont{and}
  \bibinfo{author}{\bibfnamefont{C.~G.} \bibnamefont{Van~de Walle}},
  \bibinfo{journal}{Applied Physics Letters} \textbf{\bibinfo{volume}{100}},
  \bibinfo{eid}{262104} (\bibinfo{year}{2012}).

\bibitem[{\citenamefont{Dhital et~al.}(2014)\citenamefont{Dhital, Hogan, Zhou,
  Chen, Ren, Pokharel, Okada, Heine, Tian, Yamani et~al.}}]{DhitalRuDoped}
\bibinfo{author}{\bibfnamefont{C.}~\bibnamefont{Dhital}},
  \bibinfo{author}{\bibfnamefont{T.}~\bibnamefont{Hogan}},
  \bibinfo{author}{\bibfnamefont{W.}~\bibnamefont{Zhou}},
  \bibinfo{author}{\bibfnamefont{X.}~\bibnamefont{Chen}},
  \bibinfo{author}{\bibfnamefont{Z.}~\bibnamefont{Ren}},
  \bibinfo{author}{\bibfnamefont{M.}~\bibnamefont{Pokharel}},
  \bibinfo{author}{\bibfnamefont{Y.}~\bibnamefont{Okada}},
  \bibinfo{author}{\bibfnamefont{M.}~\bibnamefont{Heine}},
  \bibinfo{author}{\bibfnamefont{W.}~\bibnamefont{Tian}},
  \bibinfo{author}{\bibfnamefont{Z.}~\bibnamefont{Yamani}},
  \bibnamefont{et~al.}, \bibinfo{journal}{Nat Commun}
  \textbf{\bibinfo{volume}{5}} (\bibinfo{year}{2014}).

\bibitem[{\citenamefont{Ikeda et~al.}(2000)\citenamefont{Ikeda, Maeno,
  Nakatsuji, Kosaka, and Uwatoko}}]{PhysRevB.62.R6089}
\bibinfo{author}{\bibfnamefont{S.-I.} \bibnamefont{Ikeda}},
  \bibinfo{author}{\bibfnamefont{Y.}~\bibnamefont{Maeno}},
  \bibinfo{author}{\bibfnamefont{S.}~\bibnamefont{Nakatsuji}},
  \bibinfo{author}{\bibfnamefont{M.}~\bibnamefont{Kosaka}}, \bibnamefont{and}
  \bibinfo{author}{\bibfnamefont{Y.}~\bibnamefont{Uwatoko}},
  \bibinfo{journal}{Phys. Rev. B} \textbf{\bibinfo{volume}{62}},
  \bibinfo{pages}{R6089} (\bibinfo{year}{2000}).

\bibitem[{\citenamefont{Nagai et~al.}(2007)\citenamefont{Nagai, Yoshida, Ikeda,
  Matsuhata, Kito, and Kosaka}}]{Nagai}
\bibinfo{author}{\bibfnamefont{I.}~\bibnamefont{Nagai}},
  \bibinfo{author}{\bibfnamefont{Y.}~\bibnamefont{Yoshida}},
  \bibinfo{author}{\bibfnamefont{S.~I.} \bibnamefont{Ikeda}},
  \bibinfo{author}{\bibfnamefont{H.}~\bibnamefont{Matsuhata}},
  \bibinfo{author}{\bibfnamefont{H.}~\bibnamefont{Kito}}, \bibnamefont{and}
  \bibinfo{author}{\bibfnamefont{M.}~\bibnamefont{Kosaka}},
  \bibinfo{journal}{Journal of Physics: Condensed Matter}
  \textbf{\bibinfo{volume}{19}}, \bibinfo{pages}{136214}
  (\bibinfo{year}{2007}).

\bibitem[{\citenamefont{Yee and Balents}(2015)}]{PhysRevX.5.021007}
\bibinfo{author}{\bibfnamefont{C.-H.} \bibnamefont{Yee}} \bibnamefont{and}
  \bibinfo{author}{\bibfnamefont{L.}~\bibnamefont{Balents}},
  \bibinfo{journal}{Phys. Rev. X} \textbf{\bibinfo{volume}{5}},
  \bibinfo{pages}{021007} (\bibinfo{year}{2015}).

\bibitem[{\citenamefont{Balzer et~al.}(2009)\citenamefont{Balzer, Kyung,
  S?n?chal, Tremblay, and Potthoff}}]{0295-5075-85-1-17002}
\bibinfo{author}{\bibfnamefont{M.}~\bibnamefont{Balzer}},
  \bibinfo{author}{\bibfnamefont{B.}~\bibnamefont{Kyung}},
  \bibinfo{author}{\bibfnamefont{D.}~\bibnamefont{S?n?chal}},
  \bibinfo{author}{\bibfnamefont{A.-M.~S.} \bibnamefont{Tremblay}},
  \bibnamefont{and} \bibinfo{author}{\bibfnamefont{M.}~\bibnamefont{Potthoff}},
  \bibinfo{journal}{EPL (Europhysics Letters)} \textbf{\bibinfo{volume}{85}},
  \bibinfo{pages}{17002} (\bibinfo{year}{2009}).

\bibitem[{\citenamefont{Zhao et~al.}(2014)\citenamefont{Zhao, Wang, Qi, Zeng,
  Hirai, Kong, Li, Park, Yuan, Jin et~al.}}]{Zhaopressure}
\bibinfo{author}{\bibfnamefont{Z.}~\bibnamefont{Zhao}},
  \bibinfo{author}{\bibfnamefont{S.}~\bibnamefont{Wang}},
  \bibinfo{author}{\bibfnamefont{T.~F.} \bibnamefont{Qi}},
  \bibinfo{author}{\bibfnamefont{Q.}~\bibnamefont{Zeng}},
  \bibinfo{author}{\bibfnamefont{S.}~\bibnamefont{Hirai}},
  \bibinfo{author}{\bibfnamefont{P.~P.} \bibnamefont{Kong}},
  \bibinfo{author}{\bibfnamefont{L.}~\bibnamefont{Li}},
  \bibinfo{author}{\bibfnamefont{C.}~\bibnamefont{Park}},
  \bibinfo{author}{\bibfnamefont{S.~J.} \bibnamefont{Yuan}},
  \bibinfo{author}{\bibfnamefont{C.~Q.} \bibnamefont{Jin}},
  \bibnamefont{et~al.}, \bibinfo{journal}{Journal of Physics: Condensed Matter}
  \textbf{\bibinfo{volume}{26}}, \bibinfo{pages}{215402}
  (\bibinfo{year}{2014}).

\end{thebibliography}
%
%
\end{document}